\documentclass{article}
\usepackage{arxiv}
\usepackage{titling}
\usepackage{titlesec}
\usepackage[T1]{fontenc}
\usepackage{authblk}

\usepackage{microtype}
\usepackage{graphicx}
\usepackage{booktabs} % for professional tables

\usepackage{hyperref}

\usepackage{amsmath}
\usepackage{algorithmic}
\usepackage{bm}
\usepackage{bbm}
\usepackage{verbatim}
\usepackage{amssymb,amsthm,amsmath}
\usepackage{dsfont}
\usepackage{mathrsfs}
\usepackage{subcaption}
\usepackage{placeins}
\usepackage{xcolor}
\usepackage{graphicx}
\usepackage[font=small,skip=5pt]{caption}
\usepackage{natbib}
\usepackage{algorithm}

\newtheorem{lemma}{Lemma}
\newtheorem{theorem}{Theorem}
\newtheorem{proposition}{Proposition}
\newtheorem{definition}{Definition}

\newcommand{\dd}{\mathrm d}
\newcommand{\mutilde}{\tilde \mu}
\def\ind{\stackrel{\mbox{\scriptsize{ind}}}{\sim}}
\def\iid{\stackrel{\mbox{\scriptsize{iid}}}{\sim}}
\newcommand{\R}{\mathbb R}
\newcommand{\E}{\mathbb E}

\usepackage[normalem]{ulem}
 
\newcommand{\MBtext}[1]{{\color{black}{#1}}}

\newcommand{\prob}{\mathrm{Pr}}
\renewcommand{\mid}{\,|\,}

\title{A Nonparametric Bayes Approach to Online Activity Prediction}

\author[1]{Mario Beraha}
\affil[1]{Department of Mathematics, Politecnico di Milano}

\author[2]{Lorenzo Masoero}
\affil[2]{Amazon.com Inc}

\author[3]{Stefano Favaro}
\affil[3]{Department of Economics and Statistics, University of Torino and Collegio Carlo Alberto}
\author[2,4]{Thomas S. Richardson}
\affil[4]{Department of Statistics, University of Washington}

\begin{document}
\maketitle

\begin{abstract}
Accurately predicting the onset of specific activities within defined timeframes holds significant importance in several applied contexts. 
In particular, accurate prediction of the number of future users that will be exposed to an intervention is an important piece of information for experimenters running online experiments (A/B tests).
In this work, we propose a novel approach to predict the number of users that will be active in a given time period, as well as the temporal trajectory needed to attain a desired user participation threshold.
We model user activity using a Bayesian nonparametric approach which allows us to capture the underlying heterogeneity in user engagement.
We derive closed-form expressions for the number of new users expected in a given period, and a simple Monte Carlo algorithm targeting the posterior distribution of the number of days needed to attain a desired number of users; the latter is important for experimental planning. 
We illustrate the performance of our approach via several experiments on synthetic and real world data, in which we show that our novel method outperforms existing competitors.
\end{abstract}

\section{Introduction}

In the context of modern digital interactions, the ability to forecast the initiation of specific activities within a given time-frame holds immense significance. Examples include the number of users who will install a software update, the number of customers who will use a new feature on a website or who will participate in an A/B test. 
Whether the focus is on estimating the number of individuals initiating an action or predicting the temporal span needed to attain a desired user participation threshold, accurate predictive models play a central role in decision making, resource allocation, and enhancing user experiences. \MBtext{See, e.g., \cite{Koh(07)} and \cite{Bak(14)} for further details on online experiments.}

\MBtext{
While participation data can be formally treated as a time series, the problem of forecasting user participation does not lend itself to time series models \citep[see][and the references therein]{richardson22a}. Moreover, intricate dynamics that underlie user engagement patterns.
}
% However, the challenge lies in the intricate dynamics that underlie user engagement patterns. 
Conventional models often assume that initiation times are identically distributed, ignoring the diverse behaviors and preferences exhibited by individuals. In reality, users demonstrate varying propensities to engage, leading to a multitude of initiation timelines. 
Recognizing this complexity, \cite{richardson22a} recently proposed a Bayesian model for the users' initiation times, which allows different behaviors to be captured, while simultaneously borrowing strength as is typical in hierarchical Bayesian models.
%However, it requires fixing the size of the population of users in advance.

In this work, we extend the framework of \cite{richardson22a} by proposing
a nonparametric Bayesian model for user activity. Contrary to classical (frequentist) nonparametric statistics, in Bayesian nonparametrics (BNP), we still assume a likelihood function belonging to a parametric family, but let the size of the parameters involved in the model be either infinite or increasing with the sample size.  
Within the context of user prediction, parameters represent the users' propensity to engage; in this work, we let the pool of potential users be unbounded. However, in any finite period, the number of active users is bounded almost surely (a.s.).
\MBtext{Therefore, contrary to \cite{richardson22a}, we do not fix the population size in advance. Beyond the mathematical elegance, this allows us to sidestep the necessity of estimating a key model parameter, either by cross-validation or based on historical data, thus reducing the modeling and computational burden.}
We study two models: the first one assumes we are given data representing daily activity for each user, while the second assumes that we are provided with the time at which each user is first included in the trial, also called the ``first triggering time''. We provide a unified treatment of these two models under the \emph{generalized Indian buffet} framework of \cite{Jam(17)}.

The central contribution of this paper is to use a \emph{scaled process} \citep{james15sp, camerlenghi2022scaled, Ber(23)}  as the prior for the infinite-dimensional parameter involved in the models. 
Compared to the completely random measure (CRM) priors that have been traditionally employed  \citep{thibaux07, titsias2007, broderick2014combinatorial, Jam(17), masoero2022more}, scaled processes offer a richer predictive structure that is well-suited to the problem of activity forecasting. \MBtext{See Section \ref{sec:crm_priors} for further discussion on the unsuitability of CRMs for forecasting tasks.}
In particular, we employ the stable Beta-scaled process \citep{camerlenghi2022scaled} that offers a sensible trade-off between analytical tractability and richness of the predictive distribution.
For the two models under examination, we provide closed-form expression for the posterior distribution of the parameters, as well as the distribution of the number of new users that will be active in a subsequent period of $D$ days, $N_D$.
To estimate how many experimental days are needed to reach the desired number of users, we consider two approaches. The first is based on the posterior distribution of such an event, which we show can be computationally burdensome in some cases. The second is a heuristic approach that aims at ``inverting'' prediction intervals for $N_D$. This requires building a global (as opposed to pointwise) prediction interval for $\{N_\ell\}_{\ell \geq 1}$ which we do by exploiting the compound Poisson representation of generalized Indian buffet processes given in \cite{Jam(17)}.

\MBtext{Our models are benchmarked on two simulation studies. Moreover, we rank the predictive performance of our models, as well as several other competitors, on 210 real world online experiments that ran on a global e-commerce platform. Our models greatly outperform competitors, ranking first more than twice as often of \cite{richardson22a}. }

\section{Two BNP Models for Activity Prediction}

We present here two models for activity prediction. The first is applicable when data consists of the daily activity of users, that is, $Z_{j, i} = 1$ if the $i$-th user is active on the $j$-th day and $0$ otherwise. 
We assume that the total possible number of users is infinite, but within any given timespan only a finite number of users are active. In this case, a useful mathematical abstraction is to represent each observation $Z_j$, $j=1,\ldots, d$ as a counting measure:
\begin{equation}\label{eq:mod1}
     Z_j := \sum_{i \geq 1} Z_{j, i} \delta_{\omega_i},
\end{equation}
where $\delta(\cdot)$ is the Dirac measure $\delta_x(A) = I_A(x)$. The measure $Z_j$ is
supported on an infinite sequence $\{\omega_i\}_{i \geq 1}$, which can be thought of as the collection of ``user-specific labels'' (e.g., anonymized user IDs) that play no other role in the subsequent analysis and simply satisfy $\text{Pr}(\omega_i \neq \omega_j) = 1$ for $i \neq j$.

% \MBnote{Connect $Y_i$ and $Z_{j,i}$?}
The second model is useful when the data consists of only the first triggering time for active users, i.e., $Y_i \in \{0, 1, \ldots, d \}$ such that $Y_i = \ell \geq 1$ if user $i$ is active for the first time on the $\ell$-th day, and $Y_i = 0$ if user $i$ is never active in the first $d$ days. 
Also in this case we assume that the potential number of users is a priori infinite, but for any timespan $d$ only a finite number of users satisfy $Y_i > 0$ and the remaining ones are not detected, i.e., $Y_i = 0$.
Again, it is useful to represent the data as a counting measure
\begin{equation}\label{eq:mod2}
    Y := \sum_{i \geq 1} Y_i \delta_{\omega_i}.
\end{equation}

We present a unified analysis of both models under the framework of \emph{trait allocations} \citep[see, e.g.,][and the references therein]{Jam(17), Cam(18)}.
In a trait allocation model, each sampling unit displays different \emph{traits} 
% $(\omega_j)_{j}$
together with a corresponding \emph{level of association}. In our setting, the sampling unit is either the day of observation $Z_j$ in \eqref{eq:mod1} or directly the entire observation period $Y$ in \eqref{eq:mod2}, with different traits corresponding to unique users, indexed by the user-specific labels. The levels of association are either the indicator that the $i$-th user is active on the $j$-th day ($Z_{j, i}$ in \eqref{eq:mod1}) or the first time the $i$--th user is active since the start of the observation period ($Y_i$ in \eqref{eq:mod2}).
We emphasize that compared to typical situations where trait allocation models have been considered \citep[see, e.g.,][and the references therein]{Jam(17), masoero2022more} here we are ``flipping'' the notion of sampling unit and trait. Indeed, usually the sampling unit corresponds to a user or subject (for instance, taking part in a genomic study) and the trait corresponds to characteristics of the individual (e.g., presence of a genomic variant). 

In a trait allocation model, it is assumed that the weights of the random measures representing the data (i.e., the $Z_{j, i}$'s in \eqref{eq:mod1} or the $Y_i$'s in \eqref{eq:mod2}) follow a common distribution $G$ and are further independent given a user-specific parameter $\tau_i$, $i \geq 1$.
User-specific parameters and user-specific labels are collected in a measure $\mutilde := \sum_{i \geq 1} \tau_i \delta_{\omega_i}$ which is the ``only'' (nonparametric) parameter in the model and upon which a prior is placed.
For notation's sake, we shall use $X = \sum_{i \geq 1} X_i \delta_{\omega_i}$ when referring to the general construction of a trait allocation, $Z_j$ when referring specifically to the case of $0-1$ valued weights as in \eqref{eq:mod1} and $Y$ when the weights take values in $\{0, \ldots, d\}$ as in \eqref{eq:mod2}.
We will use the notation $X \mid \mutilde \sim \text{TrP}(G; \mutilde)$ to denote the conditional law of a trait allocation model.

\subsection{Completely Random Measure Priors and their limitations}\label{sec:crm_priors}

Completely random measures \citep[CRMs,][]{Kin(75)} are the most widely used priors for $\tilde{\mu}$.
An a.s.\ discrete CRM over $\Omega$ is a random measure of the kind $\sum_{k \geq 1} \tau_k \delta_{\omega_k}$ such that $\{(\tau_k, \omega_k)\}_{k \geq 1}$ are the points of a Poisson point process on $\R_+ \times \Omega$. 
It is often assumed that the CRM is homogeneous, i.e., the weights $(\tau_k)_{k \geq 1}$ are independent from the atom's locations $(\omega_k)_{k \geq 1}$ and further that $\omega_k \iid P_0$, for $P_0$ a (typically diffuse) measure on $\Omega$. In this case, the law of the CRM is specified by its L\'evy measure $\nu$, which is a measure on $\R_+ \times \Omega$ of the kind $\nu(\dd s \, \dd x) = \theta \rho(s) \dd s P_0(\dd x)$ where $\theta >0$ and $\rho(s)$ is the L\'evy \emph{density}. 
Note that the measure $P_0$ governs only the user-specific labels $(\omega_k)$, and is therefore irrelevant in practice as long as it is diffuse, which ensures that different users are assigned different labels with probability one.
Therefore, for simplicity, we will fix $P_0$ to the uniform measure on $[0, 1]$ in the rest of the paper.

In the context of binary observations, notable examples of CRM priors include the Beta process \citep{Hjo(90), thibaux07}, for which $\rho(s) = s^{-1} (1 - s)^{c - 1} I_{[0, 1]}(s)$, and the three-parameter (stable) Beta process \citep{Teh(09)} that corresponds to $\rho(s) = s^{-\alpha - 1} (1 - s)^{c + \alpha - 1}  I_{[0, 1]}(s)$. In the context of integer valued observations, in addition to the Beta process \citep{Hea(16)}, the Gamma process (i.e., $\rho(s) = s^{-1} e^{-s}$) and generalized Gamma process have been considered in \cite{titsias2007} and \cite{Jam(17)}.
Note that in all the aforementioned processes, $\int_{\R_+} \rho(s) \dd s = +\infty$ so that, almost surely, the CRM has a countable number of support points. 
In our terminology, this means that the total population of potential users is infinite.
At the same time, we also have that for any $\varepsilon > 0$, $\sum_{k} I[\tau_k > \varepsilon] < +\infty$. This entails that, in any finite period, we observe a finite number of users with probability one.
However, as we allow for longer observation periods, we expect an ever increasing number of active users, diverging to $+\infty$ in the infinite-time limit.
This is a distinctive feature of Bayesian nonparametric models.
% whereby the number of relevant parameter in the model is finite for any finite dataset but allowed to grow as more data are observed.

CRM priors offer a great degree of analytical tractability, resulting in a tractable posterior, marginal, and predictive representations for any distribution $G$ and prior choice. See \cite{Jam(17), broderick2014exponential} for a detailed account.
This enables straightforward estimation of the hyperparameters of the prior via empirical Bayesian procedures \citep[see, e.g.,][]{masoero2022more}.
However, a fundamental drawback of CRM priors is that they induce a predictive distribution such that the probability of discovering ``new users'' depends only on the cardinality of the observed samples (i.e., the number of days $d$).
This could be a questionable oversimplification, possibly resulting in poor predictive performance \citep{camerlenghi2022scaled}, and motivates the study of different prior distributions not based on CRMs. 
In particular, we advocate for adopting a \emph{scaled process} as prior for $\mutilde$, showing that this choice results in richer predictive distributions while maintaining a high degree of analytical tractability.

\subsection{The Stable Beta-Scaled Process Prior}

To enrich the predictive distribution of CRM priors, \cite{camerlenghi2022scaled} proposed the so-called \emph{scaled processes} \citep{james15sp} as priors for $\tilde{\mu}$ in the context of $0$--$1$ valued data. See also \cite{Ber(23)} for an extension to the trait allocation setting.
In particular, in both papers it is shown how scaled processes derived from the $\alpha$-stable subordinator \citep{Kin(75)} provide a reasonable trade-off between analytical tractability and the ``richness'' of the predictive distribution. We recall their construction next.

First, consider an $\alpha$-stable random measure $\mu = \sum_{k \geq 1} \tau_k \delta_{\omega_k}$. That is, $\mu$ is a CRM with L\'evy intensity $\alpha s^{-1 - \alpha} \dd s P_0(\dd x)$ for $0 < \alpha < 1$.
Denote by $\Delta_1 > \Delta_2 > \cdots$ the decreasingly ordered random jumps $\tau_{k}$ of $\mu$.
Following \cite{Fer(72)} $\Delta_1$ has density $f_{\Delta_1}(\zeta) = \exp\{\alpha \int_\zeta^{+\infty} s^{-1-\alpha} \dd s\} \alpha \zeta^{-1-\alpha}$, i.e., $\Delta_{1}^{-\alpha} \sim \mathcal{E}(1)$.
Denote by $\mathcal L_\zeta(\cdot)$ the conditional distribution of $(\Delta_{k+1} / \Delta_1)_{k \geq 1}$ given $\Delta_{1} = \zeta$. 
Then, a scaled process is obtained by marginalizing $\Delta_1$ from the latter distribution. 
As noted in \cite{james15sp}, we can gain in flexibility by changing the law of $\Delta_1$, i.e., marginalizing $\mathcal L_\zeta(\cdot)$ with respect to $\zeta \sim h^*$ for any distribution $h^*$ supported on the non-negative reals.
The stable beta-scaled process is obtained by a suitable choice of $h^*$ described below.
\begin{definition}\label{def:sbsp}
    A stable beta-scaled process (SB-SP) prior is a random measure $\mutilde = \sum_{k \geq i} \tilde \tau_k \delta_{\omega_k}$ where $\omega_k \iid P_0$ and $(\tilde \tau_k)_{k \geq 1}$ is distributed as $\int \mathcal L_{\zeta}(\cdot) h_{\alpha, c, \beta}(\zeta) \dd \zeta$ where
    \[
         h_{\alpha, c, \beta}(\zeta) = \frac{\alpha \beta^{c+1}}{\Gamma(c+1)} \zeta^{- \alpha (c+1) - 1} \exp \{-\beta  \zeta^{-\alpha}\},
    \]
    for $0 < \alpha < 1$, $\beta > 0$, and $c > 0$.
    We will use the notation $\mutilde \sim \mbox{SB-SP}(\alpha, c, \beta)$.
\end{definition}

\subsection{Bayesian analysis under the SB-SP Prior}

In this section, we consider the following two models: 
\begin{equation}\label{eq:bnp1}
\begin{aligned}
    Z_j \mid \mutilde \iid \mbox{TrP}(\mathrm{Bernoulli}; \mutilde), \quad
    \mutilde  \sim \mbox{SB-SP}(\alpha, c, \beta),
\end{aligned}
\end{equation}
where the $Z_j$'s ($j=1, \ldots, d$) are random counting measures as in \eqref{eq:mod1}, and 
\begin{equation}\label{eq:bnp2}
\begin{aligned}
    Y \mid \mutilde \sim \mbox{TrP}(\mbox{TGeom}_1^d; \mutilde), \quad \mutilde \sim \mbox{SB-SP}(\alpha, c, \beta),
\end{aligned}
\end{equation}
where $Y$ is as in \eqref{eq:mod2} and $\mbox{TGeom}_d^D$ denotes a discrete distribution over $\{0, d, d+1, \ldots, D\}$ such that
\[
    \mbox{TGeom}_d^D(y; s) = \begin{cases}
        (1 - s)^{y-d} s & \text{ if } d \leq y \leq D, \\
        1 - (1 - s)^{D - d + 1} & \text{ if } y = 0.
    \end{cases}
\]
In particular, if $d=1$, $\mbox{TGeom}_d^D(y; s)$ is a \emph{truncated} geometric distribution that puts to zero all values of $y$ greater than $D$.

The two fundamental quantities that we study are the marginal distribution of the data and the posterior distribution of the random measure $\mutilde$. 
Indeed, having a tractable marginal likelihood allows us to adopt an empirical Bayesian strategy to tune the parameters of the prior in Definition \ref{def:sbsp}.
The posterior representation of $\mutilde$, instead, allows us to develop a conceptually simple Monte Carlo algorithm (and not Markov chain Monte Carlo) to address the activity prediction problem.
Below, we derive closed-form analytical expressions for both quantities of interest. These are obtained by specializing the results of \cite{camerlenghi2022scaled} and \cite{Ber(23)}.

\begin{theorem}\label{thm:postmod1}
    Let $Z_1, \ldots, Z_d$ be distributed according to \eqref{eq:bnp1}.
    Denote by $\omega^*_1, \ldots, \omega^*_{N_d}$ the observed user-specific labels, and let $M_i = \sum_{j=1}^d Z_{j, i}$.
    Then, the marginal distribution of $Z_1, \ldots, Z_d$ is
    \[
        \frac{\alpha^{N_d} \beta^{c+1}}{(\beta + \gamma_d)^{N_d + c + 1}} \frac{\Gamma(N_d + c + 1)}{\Gamma(c + 1)} \prod_{i=1}^{N_d} B(M_i - \alpha, d - M_i + 1)
    \]
    where $B(\cdot, \cdot)$ is the beta function and $\gamma_d = \alpha \sum_{i=1}^d B(1 - \alpha, i)$.
    Moreover, the posterior distribution of $\mutilde$ coincides with the law of the random measure
    \begin{equation}\label{eq:post_mu_bern}
         \sum_{i=1}^{N_d} J_i \delta_{\omega^*_i} + \mutilde^\prime,
    \end{equation}
    such that, given $\Delta_{1, c, \beta}^{-\alpha} \sim \mbox{Gamma}(N_d + c + 1, \beta + \gamma_d)$,
    \begin{itemize}
        \item $J_i \mid \Delta_{1, c, \beta} \ind \mbox{Beta}(d_i - \alpha, d - d_i + 1)$. In particular, the $J_i$'s do not depend on $\Delta_{1, c, \beta}$.
        \item $\mutilde^\prime \mid \Delta_{1, c, \beta}$ is a completely random measure with L\'evy intensity
        \begin{equation}\label{eq:levy_post}
            \Delta_{1, c, \beta}^{-\alpha} \alpha (1 - s)^d s^{-1-\alpha} I_{[0, 1]}(s) \dd s P_0(\dd x).
        \end{equation}
    \end{itemize}
\end{theorem}

In particular, note that the posterior distribution of $\mutilde$ decomposes as the sum of two parts: the first is a finite measure supported over the $N_d$ previously seen users, while the second is a random measure of the form $\mutilde^\prime = \sum_{\ell \geq 1} \tau^\prime_\ell \delta_{\omega^\prime_\ell}$. Since all the atoms (i.e., user-specific labels) are drawn from a non-atomic probability measure $P_0$, $\omega^\prime_\ell \neq \omega^*_j$ for any $\ell$ and $j$. That is, $\mutilde^\prime$ is concerned only with the potentially new users that were not active in the first $d$ days.

\begin{theorem}\label{thm:postmod2}
    Let $Y$ follow model \eqref{eq:bnp2} with the $\mbox{TGeom}_1^d$ distribution.  
    Denote by $\omega^*_1, \ldots, \omega^*_{N_d}$ the observed user-specific labels.
    Then the marginal distribution of $Y$ is
    \[
        \frac{\alpha^{N_d} \beta^{c+1}}{(\beta + \gamma_d)^{N_d + c + 1}} \frac{\Gamma(N_d + c + 1)}{\Gamma(c + 1)} \prod_{i=1}^{N_d} B(1 - \alpha, Y_i),
    \]
    where $\gamma_d$ is as in Theorem \ref{thm:postmod1}.
    Moreover, the posterior distribution of $\mutilde$ coincides with the law of the random measure
    \begin{equation}\label{eq:post_mu_geom}
        \sum_{i=1}^{N_d} J_i \delta_{\omega^*_i} + \mutilde^\prime,
    \end{equation}
    such that, given $\Delta_{1, c, \beta}^{-\alpha} \sim \mbox{Gamma}(N_d + c + 1, \beta + \gamma_d)$,
    \begin{itemize}
        \item $J_i \mid \Delta_{1, c, \beta} \sim \mbox{Beta}(1-\alpha, Y_i)$, 
        \item $\mutilde^\prime \mid \Delta_{1, c, \beta}$ is a completely random measure with L\'evy intensity as in \eqref{eq:levy_post}.
        % \[
        %     \Delta_{1, c, \beta}^{-\alpha} \alpha (1 - s)^d s^{-1-\alpha} I_{[0, 1]}(s) \dd s P_0(\dd x)
        % \]
    \end{itemize}
\end{theorem}
Surprisingly, the marginal and posterior distributions in Theorems \ref{thm:postmod1} and \ref{thm:postmod2} are very similar. In particular, the law of $\mutilde^\prime$ is identical in both settings while the law of the jumps $J_i$ associated with the previously seen users differ slightly.

\section{Activity Prediction}\label{sec:prediction}

We consider here two related problems: the first one consists in determining the follow-up timespan $D_M$ needed to reach at least $N_d + M$ active users,  while in the second one we aim predicting the number of new users in a (fixed) follow-up observational period of $D$ days.

First, observe that under both models \eqref{eq:mod1} and \eqref{eq:mod2} the first triggering time of the $\ell$-th new user follows a truncated Geometric distribution with parameter $\tilde \tau_i$ and support $\{d+1, d+2, \ldots \}$. Indeed, if the user were to trigger before day $d$, we would have observed them in the original sample.
Let $Y^\prime = \sum_{\ell \geq 1} Y^\prime_\ell \delta_{\omega^\prime_\ell}$ denote the measure tracking the first triggering time of all the unobserved users.
Denoting by $\Pi(\dd \mutilde \mid Z)$ the posterior distribution of $\mutilde$ as in Theorem \ref{thm:postmod1}, the predictive distribution of $Y^\prime$ given the sample is
\begin{align*}
    \prob(Y^\prime\mid Y) &= \int \mbox{TrP}(Y^\prime \mid \mbox{Geom}, \mutilde) \Pi(\dd \mutilde \mid Z) \\
    &= \int \mbox{TrP}(Y^\prime \mid \mbox{Geom}, \mutilde^\prime) \Pi(\dd \mutilde^\prime \mid Y)
\end{align*}
where the first equality follows from Bayes' rule, and the second equality follows by observing that the posterior of $\mutilde$ decomposes as in \eqref{eq:post_mu_geom} with $Y^\prime$ depending only on $\mutilde^\prime$.
Moreover, given that $\mutilde^\prime \mid Y \stackrel{\scriptsize d}{=} \mutilde^\prime \mid Z_1, \ldots, Z_d$ we have that the predictive rules on new users are the same under both models.

Let us start by considering the number $N_D$ of new users who will trigger in the period $\{d+1, \ldots, d + D\}$. Exploiting the compound Poisson process representation in Proposition 3.3 in \cite{Jam(17)} we obtain the following distributional characterization for $N_D$.
\begin{proposition}\label{prop:pred_nd}
    Let $N^*_\ell$ be the number of new users that are active for the first time on the $\ell$-th day ($\ell > d$). Then, given $\Delta_{1, c, \beta}$ as in Theorem \ref{thm:postmod1} we have that
    \[
        N^*_\ell \mid \Delta_{1, c, \beta} \ind \mathrm{Poisson}(\alpha \Delta_{1, c, \beta}^{-\alpha} B(1 - \alpha, \ell)) \quad \ell \geq d+1.
    \]
    Then, it follows that, letting $N_D$ be the number of users triggering in the period $\{d+1, \ldots, d + D\}$, its posterior distribution given $Y$ is
    \begin{equation}\label{eq:pred_nd}
         N_D \mid Y \sim \mathrm{NegBin}(N + c + 1, 1 - \gamma_d^D / (\beta + \gamma_0^{d + D}))
    \end{equation}
    where $\gamma_a^b = \alpha \sum_{i=1}^b B(1 - \alpha, a + i)$.
    The posterior of $N_D \mid Z$ is also equal to \eqref{eq:pred_nd}.
\end{proposition}

A straightforward way to estimate $D_M$  is to proceed by ``inverse regression'': obtain an estimator $\hat N_\ell$ of $N_\ell$ for $\ell = d+1, \ldots$ via the expected value of \eqref{eq:pred_nd} and estimate $\hat D_M$ with $\tilde D$ such that $\hat N_{\tilde D} \approx M$. 

\begin{figure}[t]
    \centering
    \includegraphics[width=0.5\textwidth]{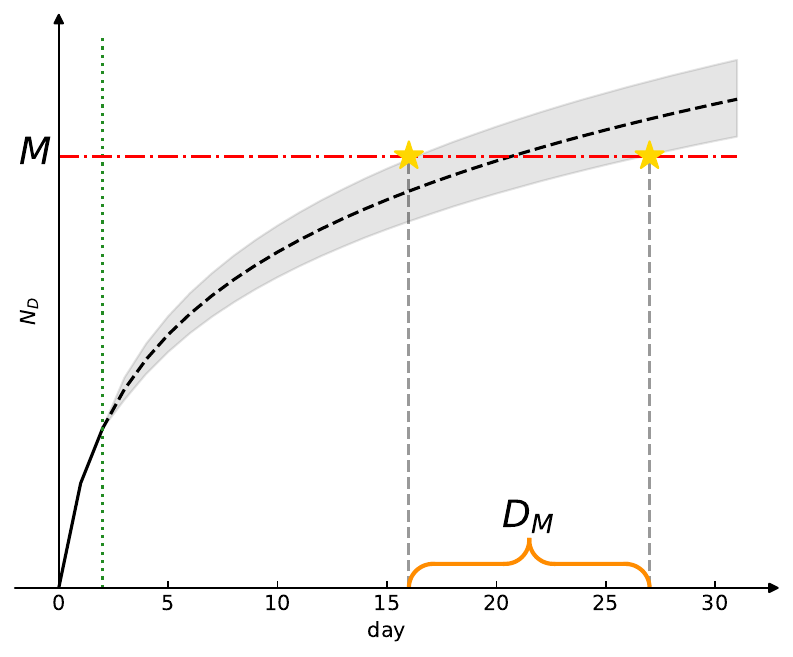}
    \caption{Inversion technique to estimate $D_M$. The solid black line represent the data. Dashed black line and gray area are the mean of \eqref{eq:pred_nd} and the global credible band. The interval for $D_M$ (orange curly bracket) is obtained by slicing the grey area at $M$.}
    \label{fig:inversion_ci}
\end{figure}

Providing prediction intervals for $D_M$ requires more care. We propose two approaches, one based on Proposition \ref{prop:pred_nd} and one based on the direct study of the posterior of $D_M$.
In the former case, we begin by constructing a global credible band, of level $\varepsilon$, of the form $\{(n^{lo}_\ell, n^{hi}_\ell)\}_{\ell = d+1}^{D_{up}}$ such that
\begin{equation}\label{eq:global_band_def}
    \text{Pr}(n^{lo}_\ell \leq N_\ell \leq n^{hi}_\ell \text{ for all } \ell \mid Y) \geq 1 - \varepsilon.
\end{equation}
Then, we slice that prediction band at $M$ to obtain an interval for $D_M$. See Figure \ref{fig:inversion_ci} for further details. 
To construct such a band, we proceed as follows. 
First we simulate $Q$ times from the posterior law of $\Delta_{1, c, \beta}$ as in Theorem \ref{thm:postmod1} and, conditional on $\Delta_{1, c, \beta}$, we sample the values $N^*_\ell$, $\ell = d+1, \ldots, D^{up}$ as in Proposition \ref{prop:pred_nd}, keeping only the $(1-\varepsilon)Q$ tuples of $(\Delta_{1, c, \beta}, \{N^*_\ell\})$ with highest density.
Then, we evaluate the $(1-\varepsilon)Q$ trajectories for $N_\ell = N_d + \sum_{j \leq \ell} N^*_j$, and set $n^{lo}_\ell$ and $n^{hi}_\ell$ equal to the minimum and maximum of the simulated values for each $\ell$.
We found that the interval for $D_M$ is robust to the chosen value of $D^{up}$ if this is sufficiently large. In our simulations, we always fix $D^{up} = 3 \hat D_M$.

The second approach we consider is to study directly the posterior of $D_M$.
Unfortunately, such a posterior has a rather complicated form. Indeed, $D_M$ can be regarded as the $M$-th order statistic of the points of a mixed Poisson process with non-constant rate, for which no closed form expression exists.
Hence, we describe next a Monte Carlo procedure that efficiently draws samples from the posterior of $D_M$.

We start by observing that to sample the first triggering times of the non-observed users $Y^\prime$, it is ``sufficient'' to first sample $\mutilde^\prime = \sum_{\ell \geq 1} \tilde\tau^\prime \delta_{\omega^\prime_\ell}$ from the corresponding posterior distribution reported in Theorems \ref{thm:postmod1} and \ref{thm:postmod2} and then, conditional on $\mutilde^\prime$, sample $Y^\prime_\ell \ind \mbox{Geom}(\tilde \tau^\prime_\ell)$ for $\ell = 1, \ldots$.
This requires truncating $\mutilde^\prime$ to a finite number (say $L$) of atoms. Simulating the largest $L$ atoms in $\mutilde^\prime$ can be achieved by exploiting the Ferguson-Klass representation of completely random measures \citep{Fer(72)}. 
See Appendix \ref{app:fk_posterior} for further details on how to simulate $\mutilde^\prime$ and an adaptive strategy to set the truncation $L$.
Then, the posterior of $D_M$ can be simulated using  Algorithm \ref{alg:post_D}.

\begin{algorithm}[t]
\caption{Posterior sampling for $D_M$}\label{alg:post_D}
\begin{algorithmic}
\STATE {\bfseries Input:} Observations $Y_1, \ldots, Y_{N_d}$, number of Monte Carlo iterations $K$.
% \KwResult{$D_1, \ldots, D_K$.}

\FOR{$k = 1, \ldots, K$}
    \STATE Sample $Y^\prime$ using Algorithm \ref{alg:post_Y_nonzero}.

    \STATE Let $Y^\prime_{(\ell)}$ be the jumps $Y^\prime_\ell$ sorted in in an increasing order.

    \STATE Set $D_k = Y^\prime_{(M - N_d)}$.
\ENDFOR
\STATE {\bfseries Return:} $D_1, \ldots, D_K$.
\end{algorithmic}
\end{algorithm}

Unfortunately, we found such an approach unsatisfactory due to the high computational cost. Indeed, sampling each jump in $\mutilde^\prime$ requires computing a non-trivial integral.
Moreover, the number of jumps simulated must be sufficiently large to ensure a negligible approximation error. 
A more efficient approach consists in simulating only those users for which $Y^\prime_\ell \in \{d + 1, \ldots, D^{up} \}$, where $D^{up}$ is a reasonable upper bound, as in the next theorem.
\begin{theorem}\label{prop:margY}
    Let $Y^\prime \mid \mutilde^\prime \sim \mathrm{TrP}(TGeom_{d+1}^{D^{up}}; \mutilde^\prime)$. Then
    \[
        Y^\prime \stackrel{\scriptsize d}{=} \sum_{\ell=1}^{K} \tilde Y^\prime_{\ell} \delta_{\omega^\prime_\ell}
    \]
    where $K \sim \mathrm{NegBin}(N + c + 1, 1 - \gamma_d^{ D^{up}} / (\beta + \gamma_0^{d + D^{up}}))$, $\omega^\prime_\ell \iid P_0$ and $\tilde Y^\prime_{\ell}$ are i.i.d. random variables supported on $\{d+1, \ldots, d + D^{up}\}$ such that
    \[
        \mathrm{Pr}(Y^\prime_{\ell} = y) \propto B(1 - \alpha, y).
    \]
\end{theorem}
The proof of Theorem \ref{prop:margY} is reported in the Appendix and makes use of the compound Poisson representation for generalized Indian buffet processes in Proposition 3.3 of \cite{Jam(17)}.
See Algorithm \ref{alg:post_Y_nonzero} for the pseudocode.

\begin{algorithm}[t]
\caption{Posterior sampling for the nonzero entries in $Y^\prime$}\label{alg:post_Y_nonzero}

\begin{algorithmic}
\STATE {\bfseries Input:} Observations $Y_1, \ldots, Y_{N_d}$.
% \KwResult{$\{Y^\prime_\ell\}_{\ell \geq 1}$.}

% Set $\phi = \sum_{j=d+1}^D B(1 - \alpha, j)$

\STATE Sample $K \sim \mathrm{NegBin}(N + c + 1, 1- \gamma_d^D / (\beta + \gamma_0^{d + D}))$

\FOR{$\ell=1 \ldots, K_\zeta$}

 \STATE Sample $Y^\prime_\ell$ from a categorical distribution over $d+1, \ldots, D$ such that
 $\text{Pr}(Y^\prime_\ell = y) \propto B(1 - \alpha, y)$

\ENDFOR
\STATE {\bfseries Return:} $\{Y^\prime_\ell\}_{\ell \geq 1}$.
\end{algorithmic}
\end{algorithm}

\section{Numerical Illustrations}

In the following, we will denote by Bernoulli model (BM) the one in \eqref{eq:bnp1} and by Geometric model (GM) the model in \eqref{eq:bnp2}.
To fit both models, we adopt an empirical Bayesian approach and estimate $\hat \alpha, \hat c, \hat \beta$ by maximizing the marginal likelihood of the data.
The negative log-marginal likelihood is not a convex function of the parameters, however we found that standard numerical optimization algorithms perform similarly in terms of the prediction errors. 
In particular, we have considered gradient descent with linesearch, BFGS, Particle Swarm, and Nelder Mead; the latter two are global optimization algorithms that do not require derivatives.
Therefore, we used BFGS as it is usually the fastest to converge.
All experiments were performed on a Macbook Pro M1 with 16GB of RAM and 8 threads.

\subsection{Comparison between the two models}\label{sec:model_comparison}

To highlight the differences between the two models, we begin by considering two simple data generating processes.
In the first (DG1) we simulate data from the Bernoulli model using the generative scheme detailed in the Appendix.
In the second (DG2) we instead start by simulating data $Y = \sum_{i=1}^{N_d} Y_i \delta_{\omega_i}$ from the Geometric model and construct the daily-activity information $Z_j$, $j=1, \ldots, d$ as follows
\[
    Z_{j, i} = \begin{cases}
        0 & \text{ if } j < Y_i \\
        1 & \text{ if } j = Y_i \\
        \varepsilon_i \mathrm{Bernoulli}((1 - \alpha) / (1 - \alpha + Y_i)) &\text{ if } j > Y_i  
    \end{cases}
\]
where $\varepsilon_i \iid \mathcal U([0, 0.5])$. That is, in DG2, after the first trigger, users tend to be less active. This resembles patterns observed, e.g., on e-commerce websites.

\begin{figure*}[t]
    \centering
    \includegraphics[width=\linewidth]{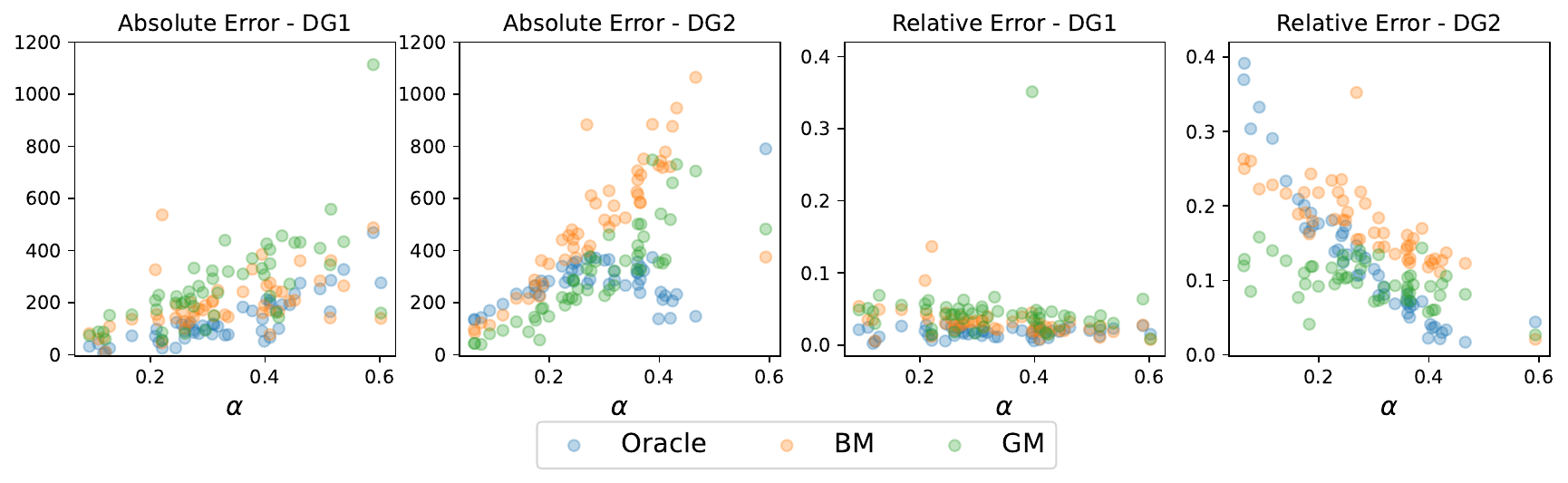}
    \caption{Absolute and relative errors for the prediction of $N_D$ in the settings described in Section \ref{sec:model_comparison}.}
    \label{fig:model_comparison_results}
\end{figure*}

For both data generating processes, we fix $d=14$, $c = 2500$, $\beta = 0.5$, and simulate 50 independent dataset that differ only in the value of $\alpha \sim \mathrm{Beta}(4, 10)$. 
\MBtext{Note that these parameters values are used only to generate the synthetic datasets, while prediction are based on estimated values $\hat \alpha, \hat c, \hat \beta$ obtained in an empirical Bayesian fashion.}
We compare the models BM and GM in a prediction task where we estimate the number of newly active users in a subsequent period of $D=14$ days by taking the expected value of the predictive distribution in \eqref{eq:pred_nd}. Since the predictive distribution is the same for both models, we are effectively comparing how well the numerical algorithms can recover the true data generating parameters when maximizing the marginal likelihoods under BM and GM. We report the absolute error $|N_D - \hat N_D|$ as well as the relative error $|N_D - \hat N_D| / N_D$ and further compare our models to a Bayesian oracle who knows the data generating parameters and uses the expected value of \eqref{eq:pred_nd} for her predictions.

Results are summarized in Figure \ref{fig:model_comparison_results}. In DG1, the oracle estimator achieves the smallest errors. BM is competitive with the oracle while GM usually yields slightly larger errors. Looking at the relative errors, we can see that these are usually less than $10\%$ and the three prediction yield comparable errors.
In contrast, under DG2, GM performs significantly better (improvements are usually greater than 10\% for the relative error) than BM across all values of $\alpha$ in the data generating process. Surprisingly, GM outperforms the oracle for smaller values of $\alpha$ as well.

In summary, it appears clear that when the data follow the assumption of BM (namely, each user is active on each given day with constant user-specific probability), BM is the best model but GM provides competitive estimates. Instead, when the data do not follow such assumptions, as in DG2, BM produces much poorer estimates than GM.
Therefore, we will focus in the subsequent analysis only on GM.

\subsection{Estimation of $D_M$}\label{sec:interval_comparison}

\begin{figure*}[t]
    \centering
    \includegraphics[width=\linewidth]{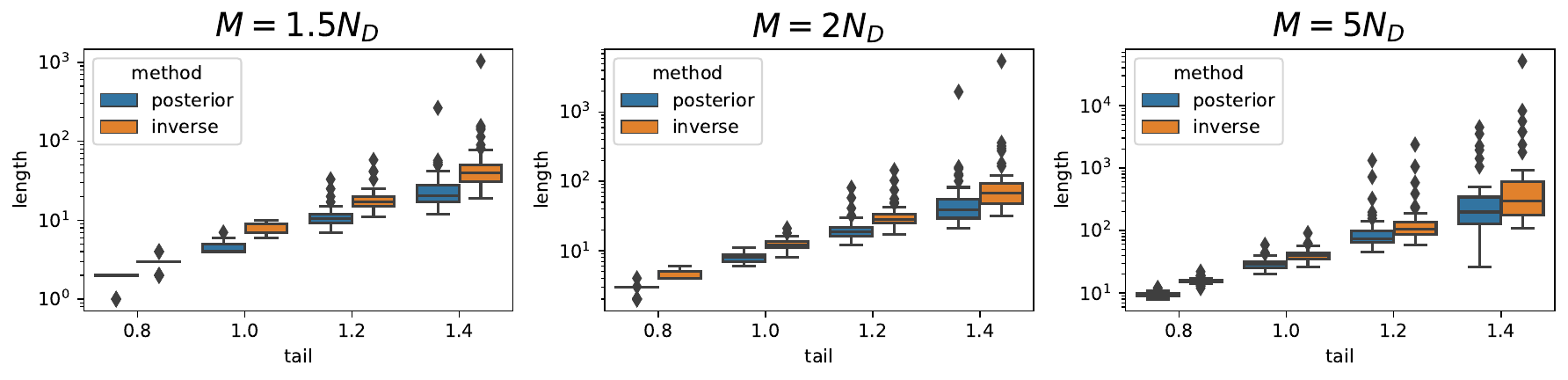}
    \caption{Length of the prediction intervals for $D_M$ for 500 simulated datasets according to the simulation in Section \ref{sec:interval_comparison}. Different plots correspond to different values of $M$. In each plot the tail parameter of the data generating process varies across the $x$-axis.}
    \label{fig:interval_lenghts}
\end{figure*}

\begin{figure*}[t]
    \centering
    \includegraphics[width=\linewidth]{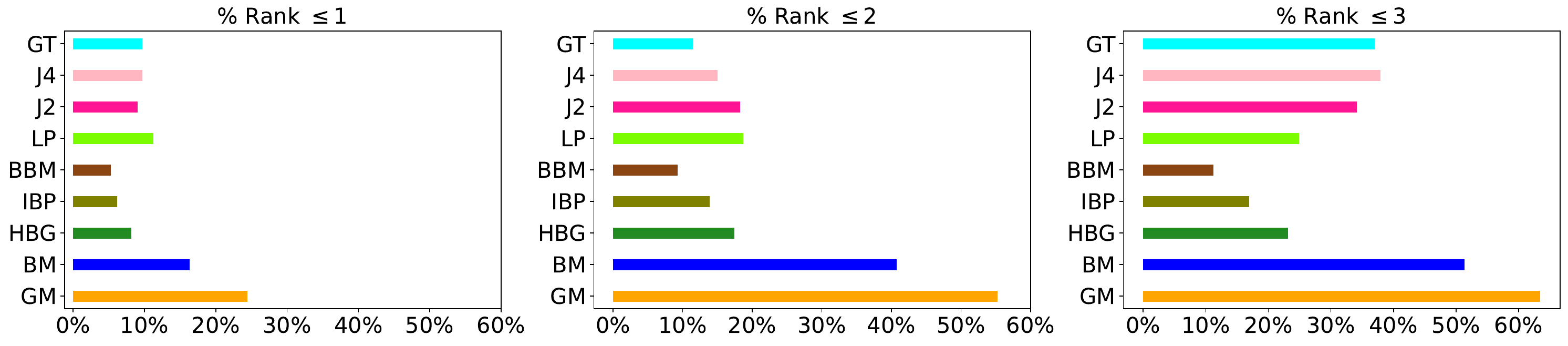}
    \caption{Ranking of the models on the 210 real datasets analyzed in Section \ref{sec:real_data}.}
    \label{fig:real_data_ranking}
\end{figure*}

We consider here the estimation of $D_M$ and compare the two approaches outlined in Section \ref{sec:prediction}.
For brevity's sake, we consider only the GM model but analogous results hold for BM.

To generate data, we assume that the total pool of possible customers is bounded by 1,000,000 and that each customer has a probability of being active in any given day of $p_i = i^{-\gamma}$, i.e., the frequency of the customers follows a Zipfian law with tail parameter $\gamma$, for $\gamma = 0.8, 1.0, 1.2, 1.4$.
We generate data for $d=14$ days, thus observing $N_d$ users, and we want to estimate the number of days needed to get $\eta N_d$ for $\eta = 1.5, 2.0, 5.0$.

We compare the estimator $\hat D_M$ and the intervals based on the inversion technique depicted in Figure \ref{fig:inversion_ci} with the posterior mean and 95\% credible intervals obtained from the posterior of $D_M$,
which is approximated using $1,000$ draws from Algorithm \ref{alg:post_D} in combination with Algorithm \ref{alg:post_Y_nonzero}.
As far as the point estimators are concerned, the two estimators give essentially the same results in terms of mean squared error. We report a comparison in the appendix. 
Focusing instead on the prediction intervals, we observe that the intervals obtained from the inversion techniques are larger than those based on the posterior of $D_M$, see Figure \ref{fig:interval_lenghts}. Moreover, the length of the intervals increases with tail parameter $\gamma$ of the data generating process. This is intuitive since for smaller values of $\gamma$ we expect to see more users in the first days while if $\gamma$ is larger it might require hundreds or thousands of days to reach the desired number of users, resulting in a greater uncertainty.
As far as the calibration of the intervals is concerned, when $M = 5N_d$ or $\gamma = 1.4$ both types of intervals are misscalibrated, their coverage varying from 60\% to 90\% depending on the setting. Instead, for $M = 1.5 N_d, 2 N_d$ and $\gamma \leq 1.0$ the posterior of $D_M$ produces (almost) perfectly calibrated intervals, while the inversion intervals have a coverage above the nominal level, as expected given that the inversion intervals are larger. Finally, when $\gamma = 1.2$, the coverage of the inversion intervals is slightly below the nominal level (around 93\%) while the posterior inference intervals have a coverage of $90\%$ (when $M = 1.5N_d$) and $85\%$ (when $M = 2N_d$).

Concerning the computational costs of producing the intervals, we note the following. To generate a draw from the posterior of $D_M$, we use Algorithm \ref{alg:post_Y_nonzero} to sample the measure $Y^\prime$. While a big-O analysis of Algorithm \ref{alg:post_Y_nonzero} eludes us, we can comment that  $K$ seems to grow linearly with the parameter $D^{up}$. 
Then one needs to sample $Y^\prime_\ell$, $\ell=1, \ldots, K$, from a discrete distribution over $\{d+1, \ldots D^{up}\}$, which requires computing $D^{up}$ probabilities. In summary, Algorithm \ref{alg:post_Y_nonzero} can be seen to scale quadratically in $D^{up}$.
Instead, the inversion approach requires computing the global credible band, that needs to sample repeatedly from the law of $N^*_\ell$, $\ell=1, \ldots, D^{up}$ in Proposition \ref{prop:pred_nd}, compute the density of the samples and evaluate the trajectories for $N_\ell$, all of which scales linearly in $D^{up}$. 
Most importantly, the computational cost of estimating the inverse interval for $D_M$ is not affected by the value of $\alpha$. On the other hand, for large values of $\alpha$, the value of $K$ in Algorithm \ref{alg:post_Y_nonzero}  can be seen to increase steeply. 
To test these insights, we performed a small simulation where we generated data from (DG2) for $d = 7$ days, with fixed parameters $\beta = 0.5$, $c=1000$ and $\alpha = 0.25, 0.5, 0.75$. We then tried to compute the inversion and posterior intervals of $D_M$ with $M = 2N_d$. Computing the inversion intervals took around $0.1$s across all simulations, while for the posterior intervals the cost ranged from $0.2$s (when $\alpha=0.25$) to almost $10$s (when $\alpha=0.75$).

% We fix the parameters of DG1 as follows: $d=14$, $c=2500$, and $\beta=0.5$ and let $\alpha = \{0.1, 0.2, \ldots, 0.9\}$. For each choice of $\alpha$ we simulate $500$ independent datasets and estimate a prediction interval for $D_M$ where $M = \gamma N_d$ for $\gamma = \{1.25, 1.5, 2.0\}$.

\subsection{Comparison on Real Datasets}\label{sec:real_data}

For our experiments on real data, we consider a collection of 210 A/B tests that ran for 28 days in production on a global e-commerce platform. For each experiment, we retain the trigger data from the first $d=7$ days as ``training'', and then assess the performance of each method by comparing the model's prediction with the realized value that occurred at $D=28$ days.

We compare a total of 9 competitors: in addition to our two models (GM and BM) we consider the hierarchical Bayesian model in \cite{richardson22a} (HGB), the Indian buffet process (IBP), the beta-Bernoulli model of \cite{ionita2009estimating} (BBM), the linear programming approach of \cite{zou2016quantifying} (LP) the jackknife estimators of order 2 and 4 (J2 and J4) \citep{burnham1979robust}, and the Good-Toulmin estimator \citep{good1953population}.
\MBtext{For all the models involving hyperparameters, these are chosen via an empirical Bayesian approach maximizing the marginal likelihood of the data for the first $d$ days.}

\MBtext{The different experiments feature an extremely diverse number of users, ranging from $10^7$ to less than $1000$. Aggregating quantitative performance metrics naively, such as reporting the mean squared error, leads to having a handful of experiments dominating the results. Likewise, as in all real data analysis, making a perfect prediction is unfeasible, especially given the small fraction of data we consider. The user participation can be greatly influenced by online advertisement campaigns, of which unfortunately we have no data. 
Therefore, we compare the different models based on a more robust metric: their top-k ranking. That is, 
} for each model we report how many times over the 210 experiments the model's prediction are the top-k best predictions for $k=1, 2, 3$.
Results are reported in Figure \ref{fig:real_data_ranking}.
We can see that our GM and BM are the best performing ones, and GM consistently outperforms BM, hinting that users preferences may vary after the first trigger. \MBtext{See the appendix for different evaluation metrics that confirm our conclusions.}

\section{Discussion}

We proposed a BNP framework to model online user's activities with a focus on predicting the number of active users in a given period of $D$ days $(N_D)$ and estimating the time needed to attain a desired level of participation of $M$ users $(D_M)$. \MBtext{Despite its central importance for experimenters, the estimation of $D_M$ has been overlooked by previous the literature.}
Our work builds on the recent contributions in \cite{camerlenghi2022scaled,Ber(23)} where the \emph{stable Beta scaled process} was proposed as a prior in trait allocation models.
\MBtext{Our work complements and extend the analysis and application of scaled process priors, showing how, compared to the use of traditional CRM prior, scaled processes yield greatly improved predictive performances without sacrificing the analytical tractability and computational convenience.}

In particular, we derived closed-form expressions for the posterior distribution of $N_D$, and proposed two strategies to estimate $D_M$.
The first, which ``inverts'' a global prediction band for $N_D$, this is computationally cheap but might produce larger intervals. The second strategy simulates directly from the posterior of $D_M$ using Monte Carlo (and not MCMC), which is slightly more demanding from the computational side.
These quantities depend crucially on the prior's parameters, for which an empirical Bayes estimate is easy thanks to the analytical tractability of our models.
We tested our approach on several examples, showing good predictive performance especially for the Geometric model, which ranked first out of 9 competitors on a large-scale experiment involving 210 real datasets collected at a global e-commerce platform.

Our Geometric model can be seen as the nonparametric version of \cite{richardson22a}. However, there is a key difference in that our model explicitly accounts (via the prior) for a power-law decay of the user-specific propensities to engage, while in \cite{richardson22a} such parameters are conditionally i.i.d. As seen in our examples, real data often supports such a hypothesis.

\bibliographystyle{chicago}
\bibliography{references}

\appendix

\section{Background Material on Bayesian nonparametrics and Random Measures}

A completely random measure (CRM) on $\Omega$ (with associated $\sigma$-algebra) is a random element $\mu$ taking values in the space of (finite) measures over $\Omega$ (with associated Borel $\sigma$-algebra) such that for any $n$ and any collection of disjoint subsets $A_1, \ldots, A_n$ of $\Omega$, the random variables $\mu(A_1), \ldots, \mu(A_n)$ are independent.
It is well-known \citep{Kin67} that a CRM can be decomposed as a sum of a deterministic measure, an atomic measure with fixed atoms and random jumps, and an atomic measure where both atoms and jumps are random and form the points of a Poisson point process.
As customary in BNP, here we discard the deterministic parts of CRMs and consider measures of the form $\mu(\cdot) = \int_{\R_+} s  N(\dd s,\, \cdot) = \sum_{k \geq 1} \tau_k \delta_{\omega_k}(\cdot)$, where $ N = \sum_{k \geq 1} \delta_{(\tau_k, \omega_k)}$ is a Poisson random measure on $\R_+ \times \Omega$ with L\'evy intensity measure $\nu(\dd s, \dd x)$.
The L\'evy intensity characterizes the distribution of $\mu$ via the L\'evy-Khintchine representation of its Laplace functional. That is, for measurable $f: \Omega \rightarrow \R_+$
\[
    \E\left[e^{-\int_{\Omega} f(x) \mu (\dd x)}\right] = \exp \left\{- \int_{\R_+ \times \Omega} (1 - e^{- s f(x)}) \nu(\dd s \, \dd x) \right\}.
\]

Our focus is on homogeneous L\'evy intensity measures, namely measures of the form $\nu(\dd s, \, \dd x) = \theta \rho(s) \dd s \, P_0(\dd x)$ where $\theta > 0$ is a parameter, $P_0$ is a nonatomic probability measure on $\Omega$ and $\rho(s) \dd s$ is a measure on $\R_+$ such that $\int_{\R_+} \rho(\dd s) = +\infty$ and $\psi(u) := \int_{\R_+}(1 - e^{-us}) \rho(s) \dd s < +\infty$ for all $u>0$.
These conditions ensure that $0 < \mu (\mathbb{W}) < +\infty$ almost surely. We write $\mu \sim \mbox{CRM}(\theta, \rho, P_0)$. 
Under a trait allocation model, the law of $\mu$ provides a natural prior distribution for the parameter of the trait process. 

We recall below the main results due to \cite{Jam(17)} for the Bayesian analysis of trait allocations under a CRM prior.
Consider the trait allocations $X_j = \sum_{k \geq 1} X_{j, k} \delta_{\omega_k}$, $j=1, \ldots, n$. Given $\mu = \sum_{k \geq 1} \tau_k \delta_{\omega_k}$ we assume that
\begin{itemize}
    \item for each $k$, $X_{j, k} \mid \mu \iid G(\cdot \mid \tau_k)$ where $G(\cdot \mid s)$ is a distribution over the nonnegative integers such that $\pi_G(s) := G(0 \mid s) > 0$
    \item the variables $X_{j, k}$ are independent across $k$, given $\mu$
    \item $\mu$ is a completely random measure with L\'evy intensity $\nu(\dd s \, \dd x) = \theta \rho(s) \dd s P_0(\dd x)$.
\end{itemize}
We use the short-hand notation
\begin{equation}\label{eq:trp_crm}
    X_j = \sum_{k \geq 1} X_{i, k} \delta_{\omega_k}\mid \mu \iid \mbox{TrP}(G; \mu), \quad \mu \sim \mbox{CRM}(\theta, \rho, P_0)  
\end{equation}
and further define
\[
    \varphi_k = \theta \int_{\R_+} \pi_G(s) (1 - \pi_G(s))^{k - 1} \rho(s) \dd s.
\]
\begin{proposition}[Marginal law, Proposition 3.1 in \cite{Jam(17)}]\label{prop:crm_marg}
    Let $X_1, \ldots, X_n$ be distributes as \eqref{eq:trp_crm}, such that the sample displays traits $\omega^*_1, \ldots, \omega^*_k$ and let $\mathcal B_j = \{i: X_i(\omega^*_j) > 0\}$, $|\mathcal B_j| = m_j$.
    Then, the marginal law of $(X_1, \ldots, X_n)$ is
    \[
        \theta^k e^{- \sum_{i=1}^n \varphi_i} \prod_{j=1}^k \int_{\R_+} (1 - \pi_G(s))^{n - m_j} \prod_{i \in \mathcal B_j} G(a_{i, j} \mid s) \rho(s) \dd s.
    \]
\end{proposition}

\begin{theorem}[Posterior law, Theorem 3.1 in \cite{Jam(17)}]\label{thm:crm_post}
    Let $X_1, \ldots, X_n$ be distributes as \eqref{eq:trp_crm}, such that the sample displays traits $\omega^*_1, \ldots, \omega^*_k$ and let $\mathcal B_j = \{i: X_i(\omega^*_j) > 0\}$, $|\mathcal B_j| = m_j$.
    Then, the posterior distribution of $\mu$ is equivalent to the distribution of
    \[
        \sum_{j=1}^k J^*_j \delta_{\omega^*_j} + \mu^\prime
    \]
    where
    \begin{enumerate}
        \item $J^*_j$ are independent positive random variables, also independent of $\mu^\prime$ with density
        \[
            f_j(s) \propto (1 - \pi_G(s))^{n - m_j} \prod_{i \in \mathcal B_j} G(a_{i, j} \mid s) \rho(s)
        \]

        \item $\mu^\prime$ is a completely random measure with L\'evy intenisty
        \[
            (1 - \pi_G(s)^{n-1} \rho(s) \dd s P_0(\dd w)
        \]
    \end{enumerate}
\end{theorem}

\begin{proposition}[Compound Poisson representation, Proposition 3.3 in \cite{Jam(17)}]\label{prop:comp_poi}
    Let $X$ be as in \eqref{eq:trp_crm}. Then the law of $X$ is equivalent to the distribution of 
    \[
        \sum_{j=1}^K \tilde A_j \delta_{\tilde \omega_k}
    \]
    where $K \sim \mbox{Poi}(\varphi_1)$, $\tilde \omega_k \iid P_0$ and the $\tilde A_k$'s are independent random variables with values in $\{1, 2, \ldots\}$ given by
    \[
        \prob(A_k = a) \propto \int G(a \mid s) \rho(s) \dd s
    \]
\end{proposition}

\section{Proofs}

We will need the following technical lemma, which follows trivially from Lemma 1 in \cite{camerlenghi2022scaled}
\begin{lemma}\label{lemma:crm}
    Let $\mutilde \sim \mbox{SB-SP}(\alpha, c, \beta)$. Then, the law of $\mutilde \mid \Delta_{1, c, \beta}$ equals the one of a CRM with L\'evy intensity
    \[
        \Delta_{1, c, \beta}^{-\alpha} \alpha s^{-1-\alpha} I_{[0, 1]}(s) \dd s P_0(\dd x)
    \]
\end{lemma}

\subsection{Proof of Theorem \ref{thm:postmod1}}

The marginal law is reported in Proposition 6 in \cite{camerlenghi2022scaled}, while the posterior distribution in Theorem 4 in \cite{camerlenghi2022scaled}.

\subsection{Proof of Theorem \ref{thm:postmod2}}

Let us start by focusing on the marginal distribution of $Y$.
By the tower property of the expectation
\[
    \prob(Y) = \E_{\mutilde}[\mathrm{TrP}(Y \mid \mutilde)] = \E_{\Delta_{1, c, \beta}} \left[ \E_{\mutilde}[\mathrm{TrP}(Y \mid \mutilde, \Delta_{1, c, \beta})] \right] 
\]
where, in the innermost expectation we recognize the marginal law of a trait allocation model endowed with a CRM prior with intensity as in Lemma \ref{lemma:crm}.
We can thus apply Proposition \ref{prop:crm_marg} (with $n=1$) to evaluate such expectation.

In particular, when $G \equiv \mbox{TGeom}_1^d$ we have $\pi_G(s) = 1 - (1 - s)^D$ and $ \varphi_1 = \sum_{i=1}^d B(1-\alpha, i)$, leading to
\begin{align*}
   &  \E_{\mutilde}[\mathrm{TrP}(Y \mid \mutilde, \Delta_{1, c, \beta})] = (\alpha \Delta_{1, c, \beta}^{-\alpha})^{N_d} e^{- \Delta_{1, c, \beta}^{-\alpha} \gamma_d} \prod_{j = 1}^{N_d} B(1 - \alpha, Y_j).
\end{align*}
The result follows by marginalizing with respect to $\Delta_{1, c, \beta}^{- \alpha} \sim \mbox{Gamma}(c, \beta)$.

As far as the posterior is concerned, we argue as above and condition on $\Delta_{1, c, \beta}$. Then an application of Theorem \ref{thm:crm_post} yields conditional representation as in the statement.
Finally, to determine the posterior law of $\Delta_{1, c, \beta}$, observe that an application of Baye's rule yields that $\Delta_{1, c, \beta} \mid Y$ has density
\[
    f_{\Delta_{1, c, \beta} \mid y}(\zeta) \propto \E_{\mutilde}[\mathrm{TrP}(Y \mid \mutilde, \Delta_{1, c, \beta})] h_{\alpha, c, \beta}(\zeta).
\]
Straightforward computations lead to recognizing the desired law, that is $\Delta_{1, c, \beta}^{- \alpha} \mid Y \sim \mbox{Gamma}(N_d + c + 1, \beta + \gamma_d)$.

\subsection{Proof of Proposition \ref{prop:pred_nd}}

Also in this case, the main idea is to first condition of $\Delta_{1, c, \beta}$ and exploit the general results in \cite{Jam(17)}. 

Let $Y^\prime = \sum_{i \geq 1} Y^\prime_i \delta_{\omega^\prime_i}$ be the first triggering times of users that are active for the first time in the days $\{d+1, \ldots, d + d^*\}$. Then, by virtue of Theorem \ref{thm:postmod2}, $Y^\prime \mid Y$ is distributed as 
\begin{equation}
    \begin{aligned}
        Y^\prime \mid \mutilde^\prime &\sim \mbox{TrP}(\mbox{TGeom}_{d+1}^{d+d^*}) \\
        \mutilde \mid \Delta_{1, c, \beta} &\sim CRM(\Delta_{1,h}^{-\alpha} \alpha (1-s)^d s^{-1-\alpha} \dd s P_0(\dd x)) \\
        \Delta_{1, c, \beta}^{-\alpha} &\sim \mbox{Gamma}(N_d + c + 1, \beta + \gamma_d).
    \end{aligned}
\end{equation}
Hence, given $\Delta_{1, c, \beta}$, $Y^\prime$ can be represented via Proposition \ref{prop:comp_poi} as
\[
    Y^\prime \mid \Delta_{1, c, \beta}, Y = \sum_{i = 1}^{N_{d^*}} \tilde Y_i \delta_{\tilde \omega_i} 
\]
where $N_{d^*} \sim \mbox{Poi}(\alpha \Delta_{1, c, \beta}^{-\alpha} \sum_{j=1}^{d^*} B(1 - \alpha, d + j))$.

Consider now $Y^{\prime \prime}$ be the triggering times of users first active on days $\{d + d^* + 1, \ldots, d + d^* + d^{**}\}$. Arguing as above it is clear that, given $\Delta_{1, c, \beta}, Y$ and $Y^\prime$,  
\[
    Y^{\prime\prime} \mid \Delta_{1, c, \beta}, Y, Y^\prime = \sum_{i = 1}^{N_{d^{**}}} \tilde Y_i \delta_{\tilde \omega_i}  
\]
where now $N_{d^{**}} \sim \mbox{Poi}(\alpha \Delta_{1, c, \beta}^{-\alpha} \sum_{j=1}^{d^{**}} B(1 - \alpha, d + d^* + j))$.
In particular, the law of $N_{d^{**}}$ does not depend on $Y^\prime$. 
Taking $d^* = d^{**} = 1$ and reasoning by induction proves verifies the statement regarding $N^*_\ell$. 

To obtain the law of $N_D$, we exploit the closeness of the Poisson family under convolution and write $N_D = \sum_{\ell = 1}^D N^*_\ell$, so that
\[
    N_D \mid \Delta_{1, c, \beta} \sim \mbox{Poi}(\alpha \Delta_{1, c, \beta}^{-\alpha} \sum_{i=1}^D B(1 - \alpha, d + i)).
\]
The results follows by integrating with respect to $\Delta_{1, c, \beta}^{- \alpha} \sim \mbox{Gamma}(N_d + c + 1, \beta + \gamma_d)$.

\subsection{Proof of Theorem \ref{prop:margY}}

The proof follows arguing as above, so we give here only a sketch. First, conditional to $\Delta_{1, c, \beta}$ we apply Proposition \ref{prop:comp_poi} which yields the representation as a counting measure over a random number of support points. 
The law of $\tilde Y^\prime_\ell$ and $\omega^\prime_\ell$ can be seen to be independent of $\Delta_{1, c, \beta}$ with the distributions reported in the statement.
Finally, $K \mid \Delta_{1, c, \beta}$ follows a Poisson distribution as in  Proposition \ref{prop:comp_poi}, so that marignally it is negative-binomial distributed with parameters as in the statement.

\section{Posterior Simulation for $Y^\prime$ via the Ferguson-Klass Algorithm}\label{app:fk_posterior}

Using the Ferguson-Klass representation of completely random measures \citep{Fer(72)} it is straightforward to sample only the $L$ largest jumps of $\mutilde^\prime = \sum_{i \geq 1} \tilde\tau^\prime_i \delta_{\omega^\prime_\ell}$.
Indeed, we have that conditionally to $\Delta_{1, c, \beta}$, $\mutilde^\prime$ is a CRM with L\'evy intensity as in \eqref{eq:levy_post}. Then, if $(\nu_i)_{i \geq 1}$ are the jump times of a standard Poisson process of unit rate (i.e., $\nu_i = \sum_{j \geq i} \varepsilon(1)$, where $\varepsilon(1)$ is an exponential random variable), we have $\tilde\tau_i = F^{-1}(\nu_i)$ where $F$ is the tail L\'evy measure
\begin{align*}
    F(v) &= \alpha \Delta_{1, c, \beta}^{-\alpha}  \int_{v}^{s} (1 - s)^d s^{-1-\alpha} \dd s \\
    &= \alpha \Delta_{1, c, \beta}^{-\alpha} B(1-v, d+1, -alpha)
\end{align*}
and $B$ is the incomplete Beta function.

The truncation level $L$ could be set deterministically, in which case we face a similar challenge as choosing the total number of users in \cite{richardson22a}, or adaptively by simulating $\tilde \tau^\prime_\ell$ until $\text{Pr}(Y_\ell  > D^{up} \mid \tilde \tau^\prime_\ell) > 1 - \delta$ for $D^{up}$ a reasonable upper bound (e.g., one year). See Algorithm \ref{alg:post_Y} for the pseudocode.

\begin{algorithm}[h!]
\caption{Posterior sampling for $Y^\prime$}\label{alg:post_Y}
\begin{algorithmic}
\STATE {\bfseries Input:} Observations $Y_1, \ldots, Y_{N_d}$. 

\STATE Sample $\zeta \sim \mbox{Gamma}(N_d + c + 1, \beta + \gamma_d)$.

\STATE Set $\ell = 1$, $E = 0$

\WHILE{Number of desired jumps is not reached}

  \STATE Set $E \leftarrow E + \mathcal E(1)$.
 
  \STATE Set $\tilde \tau_\ell$ equal to the solution of the equation
  \[
    \alpha \zeta B(1 - \tilde \tau_\ell, d + 1, - \alpha) = E.
  \]

  \STATE $\ell \leftarrow \ell + 1$

  \STATE Sample $Y^\prime_\ell \sim \mbox{Geom}(\tilde \tau_\ell) \mid Y^\prime_\ell > d$ and $\omega^\prime_\ell \sim P_0$
\ENDWHILE
\STATE {\bfseries Return:} $Y^\prime = \sum_{\ell \geq 1} Y^\prime_\ell \delta_{\omega^\prime_\ell}$.
\end{algorithmic}
\end{algorithm}

\section{Generative Schemes under the Model}

We describe below two generative schemes that can be thought of as generalizations of the IBP to our class of prior as well as to deal with non-binary scores. In particular, for the Bernoulli model this scheme is a straightforward consequence of Proposition 5 in \cite{camerlenghi2022scaled}. For the Geometric model, it is a rewriting of Theorem \ref{prop:comp_poi}.

\subsection{Bernoulli Model}

\begin{enumerate}
    \item In the first experiment day, $N_1 \sim \mbox{NegBin}(c+1, 1 - \gamma_{0}^1 / (\beta + \gamma_0?1))$ users trigger. 

    \item After $d$-days suppose we have observed $N_d$ unique users $\omega^*_1, \ldots, \omega^*_{N_d}$, such that each user has triggered at least once a day for $d_i$ days.
    Then on day $d+1$, each of the previously seen user triggers independently of each other according to a Bernoulli distribution with parameter $(d_i - \alpha) / (d - \alpha + 1)$.
    
    Moreover $N^*_{d+1} \sim \mbox{NegBin}(N_{d} + c + 1, 1 - \gamma_{d}^1 / (\beta + \gamma_0^{d+1})$ new users (i.e., previously unseen) will trigger for the first time.
\end{enumerate}

\subsection{Geometric Model}

The triggering times for users active in the period $\{1, \ldots, D^*\}$ is distributed as the random measure  in Theorem \ref{prop:comp_poi}, where we put $d = 0$ and $D^{up} = D^*$. In particular, the total number of users follows a Negative Binomial distribution with parameters $(c+1, 1 - \gamma_0^{D^*} / (\beta + \gamma_0^{D^*}))$. The triggering times are i.i.d. random variables supported on $\{1, \ldots, D^*\}$ such that 
\[
    \prob(Y_\ell = y) \propto B(1 - \alpha, y).
\]

\section{Further details for the real data analysis}

\subsection{Details on competing predictive methods}

\paragraph{HBG Model of \citep{richardson22a}}

The hierarchical Beta-Geometric model of \cite{richardson22a} assumes
\begin{equation}
    \begin{aligned}
        Y_i \mid \pi_i & \ind \text{TGeom}{1}^d, \qquad i=1, \ldots, N_{tot} \\
        \pi_i \mid \alpha, \beta & \iid \text{Beta}(\alpha, \beta) \\
        \alpha, \beta & \sim p(\alpha, \beta) \propto (\alpha + \beta)^{- 5 /2}
    \end{aligned}
\end{equation}
where $N_{tot}$ is the total population size that would be observed in an infinite time-frame.
Conditionally to $\alpha$ and $\beta$, it is possible to exploit the conjugacy of the (truncated) Geometric and Beta distributions and obtain a closed-form expression for the posterior distribution for $N_D$, that is
\[
    N_D \mid \alpha, \beta, Y \sim \text{Binomial}\left(N_{tot}, \frac{\Gamma(\beta + d + D) \Gamma(\alpha + \beta + d}{\Gamma(\alpha + \beta + d + D) \Gamma(\beta + d)}\right).
\]
We follow \cite{richardson22a} and set $N_{tot} = 10 N_d$.
The posterior distribution of $\alpha, \beta$ does not belong to a known parametric family, therefore we sample from it using Hamiltonian Monte Carlo (HMC) in \texttt{Stan}.
Compared to the rejection sampling algorithm devised in \cite{richardson22a} we found HMC to have shorter runtimes and give better predictions.

\paragraph{Indian Buffet Process}

The IBP model \citep{ibp05} assumes the same likelihood as our BM model in \eqref{eq:mod1} but places a Beta process prior on $\tilde \mu$. I.e., $\tilde \mu$ is a CRM with L\'evy intensity $s^{-1}(1 - s)^{c - 1}$ on $[0, 1]$.
The marginal and predictive laws can be easily deduced by specializing Proposition \ref{prop:crm_marg} and \ref{prop:comp_poi} to this case. 
In particular, the distribution of $N_D$ follows
\[
    N_D \sim \text{Poi}\left(\theta \sum_{j=d+1}^D \frac{c}{c + j} \right).
\]

\paragraph{Beta-Binomial Model}

The Beta-Bimonial model of \cite{ionita2009estimating} is similar to the HBG model but models the observations $Z_i$. In particular
\begin{equation}
    \begin{aligned}
        Z_{j, i} \mid \pi_i & \ind \text{Bern}{\pi_i}, \qquad i=1, \ldots, N_{d} \\
        \pi_i \mid \alpha, \beta & \iid \text{Beta}(\alpha, \beta)
    \end{aligned}
\end{equation}
In contrast to the HBG model, here $\alpha$ and $\beta$ are estimated from the data in an empirical Bayesian fashion.
See Section 1 \cite{ionita2009estimating} for further details.

\paragraph{Good-Toulmin}

We adopted the estimators proposed \citet{chakraborty2019using} for the problem of predicting the number of new genetic variants (in particular, we use the estimators provided in Equation (6) of their supplementary material). Their code is available at \url{https://github.com/c7rishi/variantprobs}.

\paragraph{Jackknife}

We consider the jackknife estimators developed by \citet{gravel2014predicting}. The $k$-th order jackknife is obtained by considering the first $k$ values of the resampling frequency spectrum; that is, by adequately weighting the number of users who appeared exactly $1, 2, \ldots, k$ times in the experiment. We adapt the code provided in \url{https://github.com/sgravel/capture_recapture/tree/master/software} to form our predictions.

\paragraph{Linear Programming}

Linear programs have been used for rare event occurrence ever since the seminal work of \citet{efron1976estimating}. Here, we adapt to our setting the predictors proposed in \citet{zou2016quantifying} via the \texttt{UnseenEST} algorithm. We adapt the implementation provided by the authors at \url{https://github.com/jameszou/unseenest} to perform our experiments.

\subsection{Further comparisons}

\begin{figure}
    \centering
    \begin{subfigure}[t]{0.5\textwidth}
        \includegraphics[width=\linewidth]{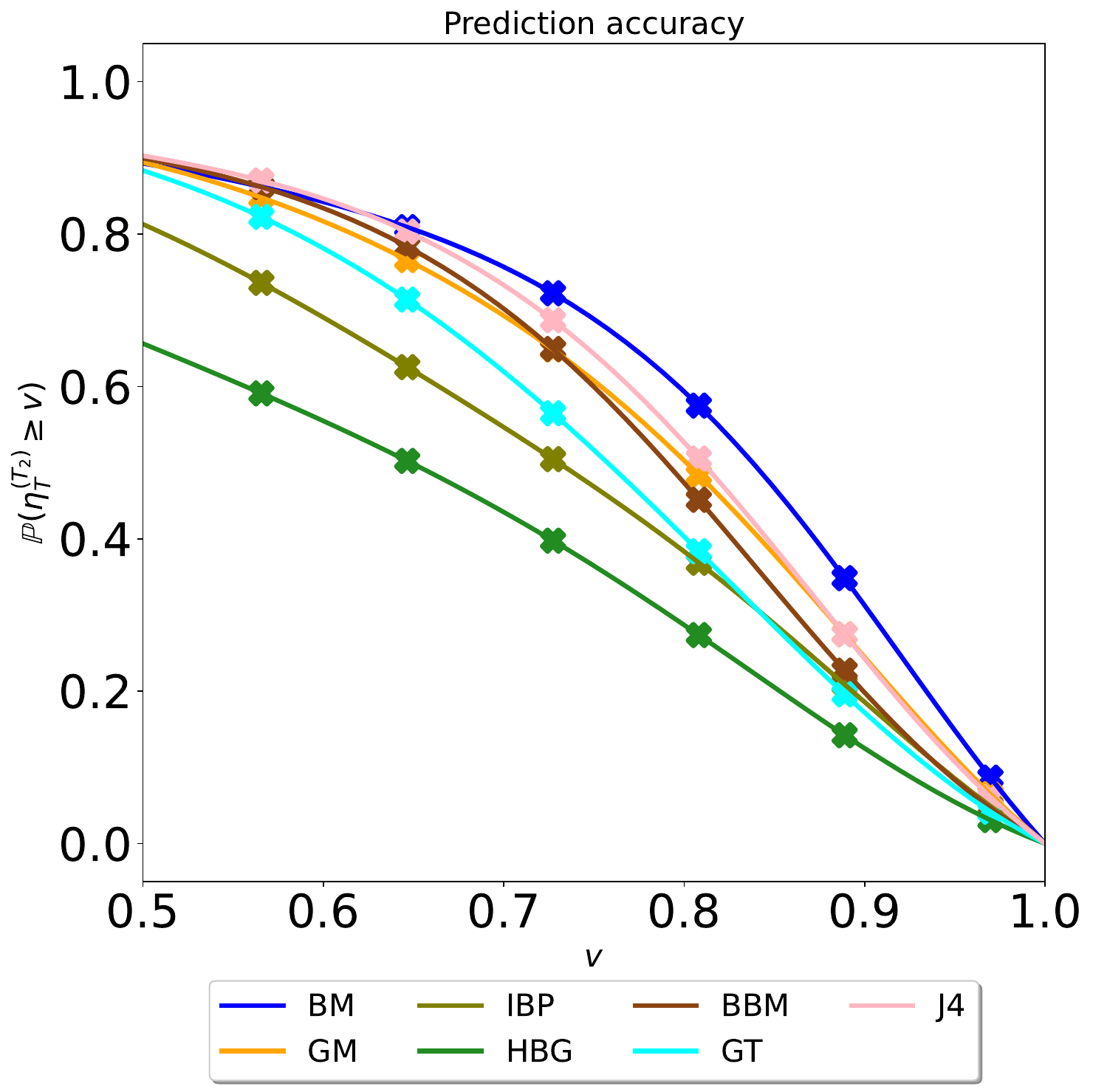}
    \end{subfigure}%
    \begin{subfigure}[t]{0.5\textwidth}
        \includegraphics[width=\linewidth]{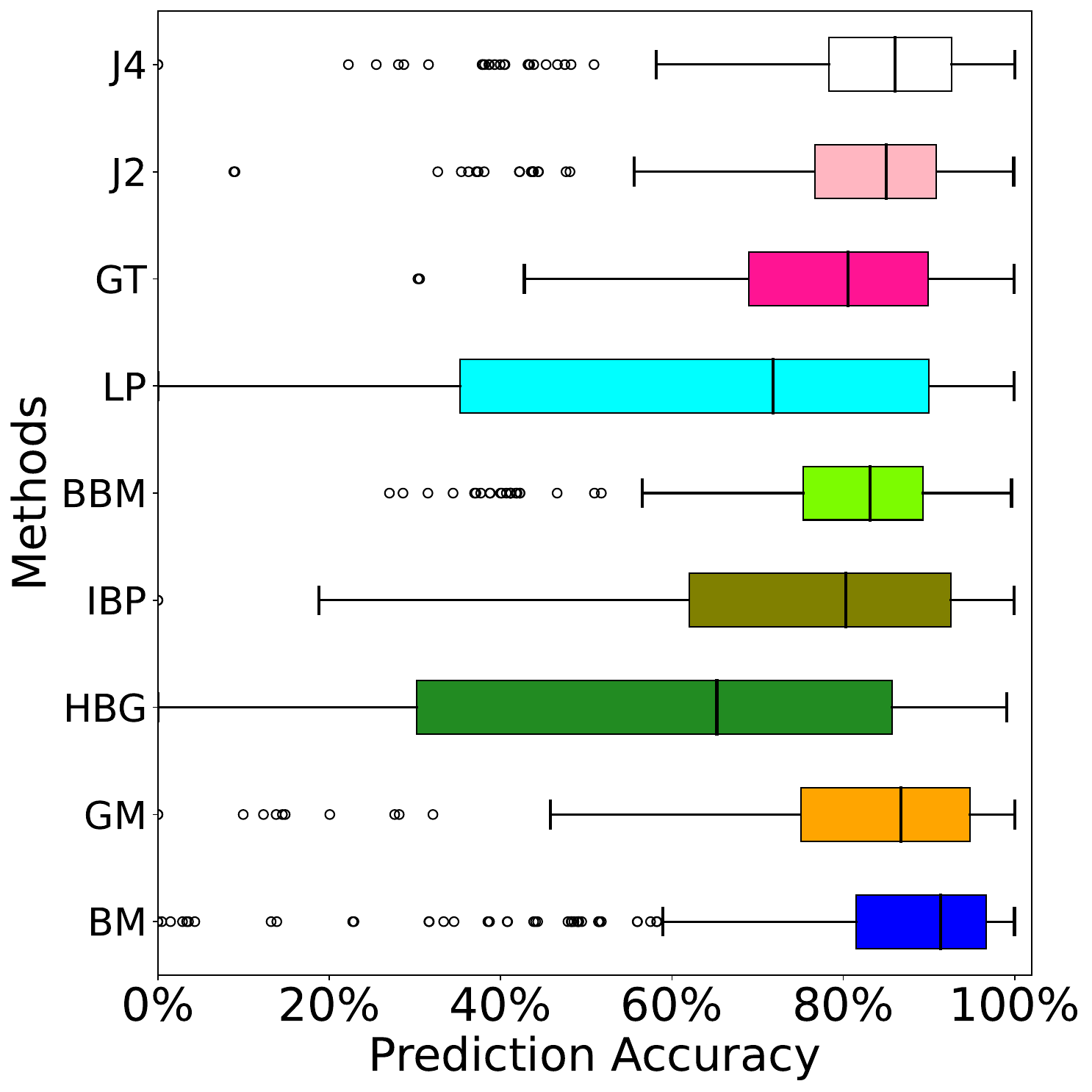}
    \end{subfigure}
    \caption{Survival function for the prediction accuracy (left) and boxplots (right) for different methods on the 210 real datasets analyzed in Section \ref{sec:real_data}.}
    \label{fig:accuracy}
\end{figure}

Let $N_D$ and $\hat N_D$ be the true and estimated number of (new) users at the last day of the experiment. We report in Figure \ref{fig:accuracy} the performance of the different models in terms of the following accuracy metric adopted from \cite{camerlenghi2022scaled}
\begin{equation*}
    \eta := 1 - \min\left\{ \frac{|N_D- \hat N_D|}{N_D}, 1\right\} \in [0,1].
\end{equation*}
This metric is equal to $1$ when the prediction is perfect and degrades to $0$ as its quality worsens.

\end{document}